\documentclass[printer]{aa}
\usepackage{txfonts}

\usepackage{graphicx}

\begin{document}

\title{Lognormal variability in BL Lacertae}

\author{Berrie Giebels \inst{1} \and Bernard Degrange \inst{1}}

\institute{
Laboratoire Leprince-Ringuet, Ecole polytechnique, CNRS/IN2P3,
 Palaiseau, France}

\date{}
 
  \abstract
   {The characterization of a time series is a powerful tool for investigating the nature
     of mechanisms that generate variability in astrophysical objects. Blazar
     variability across the entire electromagnetic spectrum is a long-standing
     puzzle, and it has been difficult to ascertain the mechanisms at play.}
   {Lognormal variability in X-ray light curves, probably related to accretion disk
    activity, has been discovered in various compact systems, such as Seyfert galaxies and X-ray
   binaries. Identifying a similar behaviour in blazars would establish a link between them.}
   {Public X-ray data from the blazar BL Lac are used to investigate the nature of its
   variability, and more precisely the flux dependency of the variability and the distribution
   of fluxes.}
   {The variations in the flux are found to have a lognormal distribution and the average amplitude of variability is
   proportional to the flux level.}
   {BL Lac is the first blazar in which lognormal X-ray variability is clearly detected. The
   light curve is orders of magnitude less variable than other blazars, with few bursting
   episodes. If this defines a specific state of the source, then the lognormality might be the
   imprint of the accretion disk on the jet, linking for the first time accretion and jet properties in a blazar.}



\maketitle

\section{Introduction}

\begin{figure}
\centering 
\includegraphics[width=9cm]{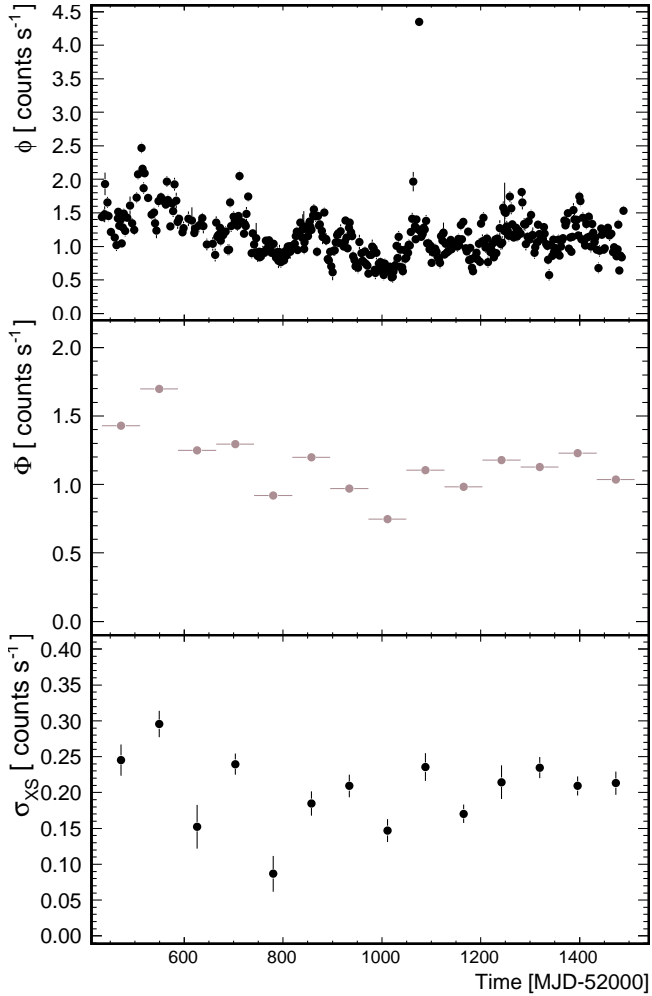}
\caption{Top panel: PCA flux light curve (in units of ${\rm counts}\,{\rm
    s^{-1}}$) of BL Lac with bins of 1 day. Middle and bottom panel: PCA mean
  flux and excess rms measured from 77 day segments (see text).}
\label{fig1}
\end{figure}

The luminosity fluctuations from astronomical sources are often used as a probe
to constrain the physical mechanisms that generate them. Nonthermal X-rays from
blazars are understood to originate in relativistic electrons cooling either through
the synchrotron or the inverse Compton mechanisms. Their spectra are usually
featureless power law continuums, with little or no thermal emission from the
disk surrounding the central black hole. If thermal
  emission is present, it is difficult to find because of the dominance of the
nonthermal radiation (\cite{per08}), complicating further the study of the link
between accretion and jet physics in these objects. Analyses have attempted to
identify in particular features in light curves where the flux varies significantly, such as peaks,
interpreted as the light crossing time through an emission zone and hence a
direct probe of its geometrical size. The cause of variability is still poorly
understood despite decades of observations across the entire electromagnetic
spectrum. 

We analyze the light curve of BL Lacertae (BL Lac), a moderately X-ray bright
blazar of the class named after itself. It has never shown dramatic
X-ray outbursts, where fluxes vary by over an order of magnitude or more, as
is often the case for this class of source when they are also very high energy
$\gamma$-ray emitters\footnote{BL Lac's spectrum extends up to $\sim 1\,{\rm
    TeV}$, see \cite{alb07}.}. The available data set has already provided a wealth of
results, the most notable being the observation of a double X-ray flare in 2005,
interpreted by \cite{mar08} as originating from a disturbance passing through two
different zones in the jet of BL Lac. The relatively major outburst-free light
curve is probed for evidence of a statistical property found in some
accreting sources, called lognormality, found in galactic as well as
extragalactic accreting sources which can be generated by a
  stationary process by taking the exponential of a Gaussian time series
  (\cite{utt05}). However, \cite{mac08} suggested 
  that variations generated within the accretion disk can modulate the
  nonthermal jet emission. Evidence of lognormal variability has also been 
  found in very high-energy gamma-ray emission from the BL Lac object PKS~2155-304
  (\cite{deg08}). Lognormal fluxes have fluctuations that are on average proportional to
the flux itself, and are the signature of an underlying multiplicative physical
process, rather than additive. This signature is particularly difficult to
find in blazars because they are relatively faint, causing the Poisson
noise to be large, and usually densely sampled only on flaring occasions and for a few
days, but even then fluxes remain modest compared to those of X-ray binaries.

\section{Data and analysis}
BL Lac has been observed routinely by the PCA detectors (\cite{jah95}) onboard
\textit{RXTE} for many years, and we analyze $6.75\times 10^5\,{\rm s}$ of
public data collected since March 2005, with exposures of a few ${\rm ks}$ per
observation. This corresponds to the \textit{RXTE} ObsID's numbered 91127, 92103
and 93135.  The PCA data were analyzed using the {\it HEASOFT 6.5.1} package
according to the guest observer facility recommended criteria, and limited to
the unit PCU2 because of known background control issues with unit
PCU0. Although the \textit{RXTE} monitoring of BL Lac did not occur on a daily
basis, a light curve is derived in 1 day bins (top panel in Fig.~\ref{fig1}),
yielding 402 flux estimates $\phi_i$ (and associated measurement errors
$\sigma_i$). In order to apply Gaussian statistics to our variance measurements,
this light curve is divided further into segments of equal length such that each
contain at least $N=20$ points (middle panel) of arithmetic mean $\Phi$. The
shortest timescale allowing this condition is found to be $T=77\,{\rm d}$, where
most of the segments have $N>20$ flux points. The excess rms $\sigma_{XS}$,
defined as $\sigma^2_{XS}= \frac{1}{N}\sum_{i=1}^{i=N} [(\phi_i - \Phi)^2 -
  \sigma^2_i]$ (\cite{vau03}), which estimates the rms corrected for the Poisson
noise, is then derived for each of the segments (bottom panel).  Varying this
length $T$, or requiring a larger amount of points $N$ in the variance
estimation, does not change significantly the results presented here. The error
in the excess variance is the 68\% confidence level estimated from $10^4$
simulated light curves, where a random flux from a Gaussian of dispersion equal
to the flux error $\sigma_i$ is added to each individual flux estimate $\phi_i$
(as in Appendix B in \cite{vau03}). Its relatively flat light curve compared to
other X-ray blazars, along with the highest flux point visible at MJD 53075,
where the flux more than tripled, indicates that the source spends most of its
time in a nonflaring state, which we refer to as its ``quiescent'' state. The
short forementioned flare is clearly an exception in the overall variability of
this source. A more detailed analysis of this single $2.4\,{\rm ks}$
observation shows that the 2--10 keV flux of $5.2\times 10^{-11}\,{\rm erg}\,{\rm
  cm^{-2}}\,{\rm s^{-1}}$ was associated with a spectral power law index of
$\Gamma=2.5\pm0.1$, much brighter than the $1-3\times 10^{-11}\,{\rm erg}\,{\rm
  cm^{-2}}\,{\rm s^{-1}}$ fluxes reported in Fig.~2 of \cite{mar08}, but also
very soft compared to its brightness. Since our aim is to characterize the
long-term lightcurve, this single observation is excluded from the analysis,
removing one flux estimation out of 402. It becomes obvious why
removing this event is also conservative in terms of establishing the nature
of the variability.

The light curve fluxes are presented in a histogram in Fig.~\ref{fig2} (top panel),
along with the result of a chi-square Gaussian fit. The fit of the distribution
of the fluxes has a chi-square probability of $p(\chi^2)=0.015$ ($\chi^2=26.4$,
${\rm n.d.f}=13$), and a width of $\sigma=0.26\,{\rm s^{-1}}$, indicating that
statistical fluctuations alone cannot explain the fit. The natural logarithm of
the fluxes provide a much better Gaussian fit with $p(\chi^2)=0.47$
($\chi^2=12.75$, ${\rm n.d.f.}=13$) (bottom panel), and a width of
$\sigma=0.26$. The scatter plot of $\sigma_{XS}$ versus the average flux in
Fig.~\ref{fig3}, in which the short flare is excluded, shows a clear correlation
($\rho=0.73$). A fit to a line ($\chi^2=33$, ${\rm n.d.f}=12$) yields
$\overline{\sigma_{XS}} \propto (0.15\pm 0.02)\Phi$. An F-test shows that the
improvement to a line fit is significant at a confidence level exceeding $99\%$
compared to a constant fit ($\chi^2=85$, ${\rm n.d.f}=13$). Including the flare
observation causes the linear fit to be further preferred to the constant fit,
and increases the slope of the linear fit by $\sim 4\%$. It is important to
remark that the flare observation is statistically very unlikely to be just
a large variation of the same lognormal process, which is another indication
that it is more likely to be a physically distinct state.

\section{Discussion}

\begin{figure}
\centering 
\includegraphics[width=10cm]{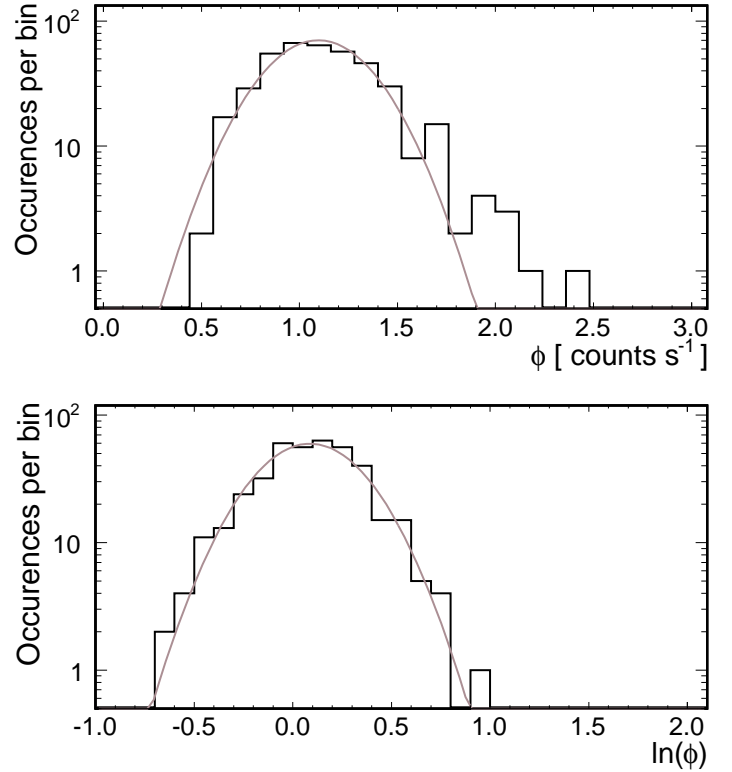}
\caption{Distribution of the fluxes for the considered period (top) and the
  distribution of the logarithm of the fluxes (bottom). The lines are the result
  of a Gaussian fit to the data. The improvement to the fit provided by the
  logarithmic function is clearly apparent in that it
  reproduces more successfully the steeper-than-Gaussian left-hand part of the distribution, as well as the
  broader, tail-like right hand part, which are characteristic of the lognormal
  distribution. The single flux estimation of the flare on MJD 54075 has been
  excluded so as not to favour the logarithmic fit. }
\label{fig2}
\end{figure}

\begin{figure}
\centering 
\includegraphics[width=9cm]{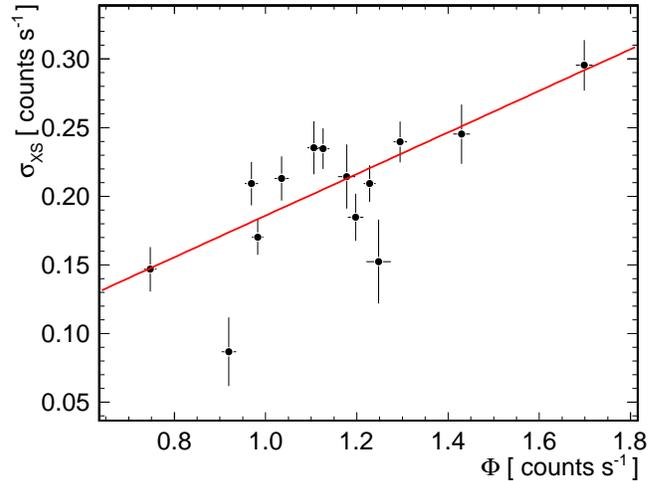}
\caption{Scatter plot of the excess variance versus the average of the fluxes for which the
  excess was determined. The line is a linear fit showing $\sigma_{XS} \propto (0.15\pm
  0.02)\,\Phi$.}
\label{fig3}
\end{figure}

The lognormality of the flux distributions, and the evidence that the
fluctuations in the flux are proportional to the flux, imply that the
variations are lognormal (\cite{ait63}). The proportionality factor of 0.15 in
Fig.~\ref{fig3} is comparable to that (0.17 with a 1-day mean flux binning
using \textit{ROSAT} data, and 0.26 with a 0.5-day binning using \textit{ASCA}
data) found in the most X-ray variable narrow-line Seyfert 1 galaxy IRAS
13244$-$3809 by \cite{gas04}, the only other known case of an extragalactic
source exhibiting lognormal X-ray flux variations. The width in Fig.~\ref{fig2}
of $\sigma=0.26$ is however smaller ($0.44\,{\rm dex}$ and $0.54\,{\rm dex}$ for
the two different data sets mentioned above) than in IRAS 13244$-$3809, meaning
that the fluctuations are smaller in this BL Lac than in the Seyfert galaxy. A
linear correlation between the excess rms and the average flux was also found
in the Seyfert 1 galaxy Mrk 766 (\cite{vau03}). The physical process responsible
for X-ray emission in Seyfert 1 galaxies is probably thermal emission from the
accretion disk. We also note that a linear flux-rms
relationship was found for Galactic accreting sources such as X-ray binaries
Cyg X-1 and SAX J 1808.4$-$3658 by \cite{utt01}, of proportionality factors of
0.33 and 0.31, respectively. A specific observation of Cyg X-1 exhibited a
remarkable lognormal fit to its X-ray flux distribution (\cite{utt05}). The
origin of the linear flux-rms was interpreted by \cite{utt01} as originating from the
subdivision of magnetic reconnection energy release as an avalanche occuring on
large scales in the corona. The same authors note that the radius-dependent
fluctuations in the mass accretion rate in the model of \cite{lyu97} can also
explain this relationship. It seems possible that the photon breeding mechanism
(\cite{ste08}, and references therein), which generates highly non-linear light
curves in relativistic jets and operates in sites that are not necessarily
directly connected to accretion properties, could be an alternative explanation
of lognormal blazar variability, but this remains to be investigated since the
statistical properties of the photon luminosity from this model are not known at
this point.

These striking similarities suggest that the variability mechanism causing the X-ray modulation
in BL Lac, outside of flaring events, is independent of the emission mechanism,
and originates in the disk, rather than in the jet, provided these lognormal fluctuations are indeed
present in the disk. The nonthermal emission from BL Lac is usually modeled by the
synchrotron self-Compton scenario (SSC, see e.g., \cite{mar92,bic02}), where relativistic
electrons cool either via the synchrotron or the inverse Compton mechanism. The comptonized
photons are assumed to be predominantly the synchrotron photons themselves, rather than photons
coming from the outside the emission region as in the external Compton model (EC, see
e.g. \cite{sik94}), given the absence of thermal signatures in the emission spectrum of BL
Lac. However, these mechanisms do not provide a way to generate lognormal variability if one of
the parameters affecting the predicted flux (e.g., magnetic field, Doppler factor, injection rate)
does not have a lognormal time-dependence itself. If fluctuations in the disk
are indeed the origin of the lognormal radiation in BL Lac, then the most
appropriate hypothesis would be one in which lognormal fluctuations in the accretion rate yield an injection rate with similar
properties. If however external photons are at least partly responsible for the cooling of the X-ray emitting
electrons, then lognormal fluctuations in the disk luminosity would have a similar result. An
ambitious observation programme targeting other X-ray bright blazars, over similar time scales, and
with a similar sampling quality, would allow us to search for lognormal
variability, and perhaps solve an important piece in the puzzle about blazar variability.

\section*{Acknowledgements}

We thank the anonymous referee for the careful reading and the constructive remarks, which have improved
this report. BG gratefully acknowledges comments from D. Barret and G. Dubus on
an early draft.

\end{document}